\documentclass[twocolumn,amsmath,aps,showpacs,a4paper]{revtex4}

\usepackage{graphicx}
\newcommand{\dif}{\mathrm{d}}

\newcommand{\vv}{\mathbf{v}}
\newcommand{\uu}{\hat{\mathbf{u}}}
\newcommand{\nn}{\hat{\mathbf{n}}}
\newcommand{\FF}{\mathbf{F}}
\newcommand{\Fc}{\mathbf{F}^c}
\newcommand{\TT}{\mathbf{T}}
\newcommand{\Tc}{\mathbf{T}^c}
\newcommand{\ct}{\cos \theta}
\newcommand{\st}{\sin \theta}

\begin{document}

\title{Translational and rotational friction on a colloidal rod near a wall}
\author{J.\ T.\ Padding}
\affiliation{Computational Biophysics, University of Twente,
           P.O. Box 217, 7500 AE  Enschede, The Netherlands}
\author{W.\ J.\ Briels}
\affiliation{Computational Biophysics, University of Twente,
           P.O. Box 217, 7500 AE  Enschede, The Netherlands}
\date{\today}

\begin{abstract}
We present particulate simulation results for translational and rotational friction components of a shish-kebab model of a colloidal
rod with aspect ratio (length over diameter) $L/D = 10$ in the presence of a planar hard wall.
Hydrodynamic interactions between rod and wall cause an overall enhancement of the friction tensor components. We find that
the friction enhancements to reasonable approximation scale inversely linear with the closest distance $d$ between the rod surface
and the wall, for $d$ in the range between $D/8$ and $L$.
The dependence of the wall-induced friction on the angle $\theta$ between the long axis of the rod and the normal to the wall
is studied and fitted with simple polynomials in $\cos \theta$.
\end{abstract}


\maketitle

\section{Introduction}

A particle suspended in a fluid, moving in the vicinity of a stationary wall, feels a viscous drag force which is larger than the
viscous drag it would experience in the bulk fluid \cite{Faxen,Brenner,Brenner1962}.
This may intuitively be understood by considering the special case of a particle moving towards (away from)
a wall: fluid needs to be squeezed out (sucked into) the gap between the particle and the wall. Even when the particle
is relatively far away from the wall the hindering effects of the wall are still felt through the long-ranged 
hydrodynamic inteactions. This has important consequences for practical applications where flow and time are issues. Especially
for microfluidic applications \cite{Squires2005}, where large surface to volume ratios are encountered, it is important to
understand the fundamentals of near-wall dynamics.

When dealing with colloidal particles random forces should also be taken into
account \cite{Dhont}. The random forces are caused by temporary imbalances in the collisions with the solvent molecules, and lead to 
diffusive (Brownian) motion of the colloidal particles. The diffusive behaviour of nanometer to micrometer sized particles near
walls is essential for the transient kinetics of phenomena such as wetting and particle deposition on a substrate \cite{Brady2009}.

The (anisotropic) diffusion tensor $\mathbf{D}$ of a colloidal particle is related to the anisotropic friction tensor $\mathbf{\Xi}$ by the generalised
Einstein relation $\mathbf{D} = k_BT \mathbf{\Xi}^{-1}$, where $k_BT$ is the thermal energy.
The friction tensor in the presence of a stick boundary wall is difficult to obtain theoretically. Analytical expressions 
in the creeping flow limit (applicable to colloidal particles) are known, but are limited to the case of a spherical particle \cite{Brenner,Goldman}
or to a non-spherical particle whose major (hydrodynamic) axes are aligned with the wall and which is
far removed from the wall \cite{Brenner1962,CoxBrenner}.
In the general case, particles are not aligned with the wall and/or may not be far removed from it. One then has to resort to experiment or
numerical evaluation to obtain the friction or diffusion tensor.

Experimentally, optical microscopy \cite{Kihm,Banerjee,Carbajal,Zahn1994},
total internal reflection microscopy (TIRM) \cite{BevanPrieve,HuangBreuer},
and evanescent wave dynamic light scattering (EWDLS) \cite{Brady2009,HolmqvistPRE,HolmqvistJCP}
have been used to determine the diffusivity of particles near a wall. The latter two experimental techniques use the
short penetration depth of an evanescent wave under total internal reflection conditions,
where in EWDLS this is combined with dynamic light scattering. In EWDLS the
different components of the diffusion tensor
may be obtained from the intensity time-autocorrelation, but this requires several careful theoretical interpretations \cite{HolmqvistJCP}.

Numerical evaluation of the friction on a particle can be performed in several ways: by numerical summation of the forces due to a large number of Stokeslets
distributed over the walls and surfaces of the particles \cite{Dhont,Durlofsky1989}, possibly including image singularities to efficiently
capture the effect of a planar wall \cite{Bossis1991,Meunier1994,SwanBrady}, or by a multipole expansion of the force densities induced on the spheres, also with an
image representation to account for a planar wall \cite{Cichocki,Ekiel-Jezewska2008}.

In this paper we will present an alternative way to determine the friction on a colloidal particle, using molecular dynamics simulations which
explicitly include the solvent particles. Because of the large difference in length scales between a colloidal particle and a solvent molecule,
it is impossible to perform such simulations in full atomistic detail. Some form of coarse-graining is necessary. Here we choose the 
Stochastic Rotation Dynamics (SRD) method to effectively represent the solvent \cite{Malevanets}. The solvent interacts with walls and
colloidal particles through excluded volume interactions \cite{Malevanets00,Padding05,Padding06}. Using this approach, we determine
the friction on a shish-kebab model of a rod of aspect ratio 10, i.e. 10 touching spheres on a straight line, as
a function of distance to and angle ($\theta$) with a planar wall. We will show that the functional dependence of the wall-induced friction enhancement 
can be reasonably well described by the inverse of the closest distance $d$ between rod surface and the wall (for $d$ in the range between one eighth the
rod diameter and the rod length), and an angular dependence which
may be expressed as a simple polynomial in $\cos \theta$. These results serve as a first example to show how SRD simulations may
be used to determine the friction on particles of non-trivial shape in a non-trivial orientation with respect to confining boundaries.

Stokesian dynamics codes, such as those developed by Brady and co-workers \cite{SwanBrady} and Cichocki
and co-workers \cite{Cichocki,Ekiel-Jezewska2008}, have other sophisticated methods to
determine the friction tensor. The results from such methods are generally more precise
(if sufficient care is taken in its implementation and choice of parameters) than those from SRD simulations, because in SRD the solvent dynamics is stochastic and the resolution seems to be limited by the collision cell size. However, as we will show, in practice
the resolution is better than a collision cell size, and the influence of stochasticy can be severely reduced by
taking long time averages, leading to an acceptable precision for making predictions.
The biggest advantage of SRD over Stokesian dynamics is the extreme simplicity of the implementation
of the SRD method when non-trivial shapes and complex confining boundaries are involved.
In Stokesian dynamics methods one has to deal with complicated multipole expansions and image representations.
If the embedded particle shape is non-trivial, or if the confining boundaries are not planar but more complex,
they have to be represented by assemblies of different sized spheres. In contrast, in SRD, being essentially a
molecular dynamics technique with a simple additional rule for momentum exchange, all that is needed is a rule to
determine when a solvent point particle overlaps with the embedded particle or wall. The SRD method then
automatically takes care of all hydrodynamic interactions between complex shapes.

This paper is organised as follows. In section \ref{sec_method} we give details of the simulations method and the constraint technique
by which we determine the hydrodynamic friction. In section \ref{sec_sphere} we validate the method by comparing simulations
of a sphere near a planar wall with known analytical expressions. Then in section \ref{sec_rod} we study the friction on a rod.
In section \ref{sec_concl} we conclude.

\section{Method}
\label{sec_method}

\subsection{Simulation details}

In Stochastic Rotation Dynamics (SRD) \cite{Malevanets} a fluid is represented by $N_f$ ideal particles of mass $m$. After propagating the
particles for a time $\delta t_c$, the system is partitioned in cubic cells of volume $a_0^3$. The velocities relative to the
centre-of-mass velocity of each separate cell are rotated over a fixed angle $\alpha$ around a random axis. This procedure
conserves mass, momentum and energy, and yields the correct hydrodynamic (Navier-Stokes) equations, including the effect
of thermal noise \cite{Malevanets}. The fluid particles only interact with each other through the rotation procedure, which can be
viewed as a coarse-graining of particle collisions over time and space.

To simulate the colloidal spheres, we follow our earlier implementation described in \cite{Padding05}.
Throughout this paper our results are described in units of SRD mass $m$, SRD cell size $a_0$
and thermal energy $k_BT$. The number density (average number of SRD particles per SRD cell)
is fixed at $\gamma = 5$, the rotation angle is $\alpha = \pi/2$, and the collision interval $\delta t_c = 0.1 t_0$,
with time units $t_0 = a_0(m/k_BT )^{1/2}$; this corresponds to a mean-free path of $\lambda_{\mathrm{free}} \approx 0.1a_0$.
In our units these choices mean that the fluid viscosity takes the value $\eta = 2.5 m/a_0t_0$ and the kinematic
viscosity is $\nu = 0.5 a_0^2/t_0$. The Schmidt number Sc, which measures the rate of momentum
(vorticity) diffusion relative to the rate of mass transfer, is given by $\mathrm{Sc} = \nu/D_f \approx 5$, where $D_f$
is the fluid self-diffusion constant \cite{Kikuchi04,Ihle04,Padding06}. In a gas $\mathrm{Sc} \sim 1$, momentum is mainly transported by
moving particles, whereas in a liquid Sc is much larger and momentum is primarily transported
by interparticle collisions. For our purposes it is only important that vorticity diffuses faster
than the particles do.

Stochastic stick boundaries are implemented as described in Ref. \cite{Padding05}. In short, SRD particles which overlap with a wall or sphere
are bounced back into the solvent with tangential and normal velocities from a thermal distribution. The change of momentum is used to
calculate the force on the boundary.
We note that despite the fact that the boundaries are taken into account through stochastic collision
rules, the \emph{average} effect is that of a classical stick boundary as often employed in
(Stokesian) continuum mechanics.
Because we will determine frictions by taking long time averages, the average flow velocities close to the boundaries will be effectively zero in all directions.
In this work we set the sphere diameter to $D = 8 a_0$, which is
sufficiently large to accurately resolve the hydrodynamic field to distances as small as $D/16$, as already shown in Refs. \cite{Padding05,Padding06}.

The method will be validated here again by comparing the friction between a sphere and a wall with known expressions from hydrodynamic
theory.

Walls are present at $z = 0$ and $z = L_z$, i.e. the wall normal $\nn$ is in the $z$-direction.
Simulations containing a single sphere are performed in a box of dimensions $L_x = L_y = L_z = 80 a_0$, corresponding to $10 D$
in each direction. Simulations containing a rod with its longest axis along $\nn$ are performed in a box of $10 D \times 10 D \times 20 D$.
All other rod simulations are performed in a box of dimensions $20 D \times 10 D \times 20 D$. The latter boxes contain 
approximately $10^7$ SRD particles.
Note that the finite box dimensions imply that there will still be an important self-interaction of the rod with its periodic images. Larger boxes
are computationally too expensive. In this work we will assume that the dependence of the hydrodynamic friction on
angle and distance to the wall is dominated by the presence of the wall itself and that self-interactions with periodic images
lead to a simple overall multiplication factor of the friction.

All simulations were run for a time of $5 \times 10^5\, t_0$, i.e. 5 million collision time steps. In CPU time this corresponds to about 4 weeks on a
modern single core processor. Such long run times are necessary to attain sufficient accuracy in the determination of the friction, the method of
which is explained in the following paragraph.

\subsection{Determination of the friction}

The translational friction tensor $\mathbf{\Xi}$ transforms the translational velocity $\vv$ of the centre-of-mass of an object to the 
friction force $\FF$ it experiences. Similarly, the rotational friction tensor $\mathbf{Z}$ transforms the rotational velocity $\mathbf{\omega}$
around the centre-of-mass to the friction torque $\TT$. In formula:
\begin{eqnarray}
\FF &=& -\mathbf{\Xi} \cdot \vv \\
\TT &=& -\mathbf{Z} \cdot \mathbf{\omega}.
\end{eqnarray}
In our simulations, the translational friction tensor is determined without actually moving the rod by measuring the time correlation of the
constraint force $\Fc$ needed to keep the rod at a fixed position \cite{Akkermans,Padding}:
\begin{equation}
\Xi_{\alpha \beta} = \frac{1}{k_BT} \int_0^{\infty} \dif \tau \left\langle F^c_{\alpha}(t_0 + \tau) F^c_{\beta}(t_0) \right\rangle_{t_0}, \label{eq_fricintegral}
\end{equation}
where $\alpha,\beta \in \left\{x,y,z\right\}$ and the subscript to the pointy brackets indicates an average over many time origins $t_0$.
Similarly, the rotational friction tensor is determined from the time correlation of the constraint torque $\Tc$ needed to keep
the rod at a fixed orientation:
\begin{equation}
Z_{\alpha \beta} = \frac{1}{k_BT} \int_0^{\infty} \dif \tau \left\langle T^c_{\alpha}(t_0 + \tau) T^c_{\beta}(t_0) \right\rangle_{t_0}. \label{eq_rotfricintegral}
\end{equation}

It is important to realise that there are two sources of friction on a large particle \cite{Malevanets00}. The first comes from the hydrodynamic dissipation
in the fluid induced by motion of the particle and may be obtained by solving Stokes' equation, integrating the solvent stress over the surface of the particle.
For translation of a sphere of diameter $D$ in an infinite fluid this yields the familiar (isotropic) friction $\xi_h = 3 \pi \eta D$.
The second contribution is the Enskog friction \cite{Hynes77,Hynes79}, which is the friction that a large particle would experience if it were dragged
through a non-hydrodynamic ideal gas, \textit{i.e.} through a gas where the velocities of the particles are uncorrelated in space and time and distributed
according to the Maxwell-Boltzmann law. For a heavy sphere with diffusing boundaries (which randomly scatter colliding particles according to the
Maxwell-Boltzmann law) the translational Enskog friction is given by
$\xi_e = \frac23 \sqrt{2\pi m k_BT} \gamma D^2 (1+2\chi)/(1+\chi)$, where $\chi = 2/5$ is the gyration ratio.
It has been confirmed \cite{Padding05,Padding06,Hynes77,Hynes79,LeeKapral04} that the two sources of friction act in parallel, \textit{i.e.} that the total
friction $\xi$ is given by
\begin{equation}
\frac{1}{\xi} = \frac{1}{\xi_h} + \frac{1}{\xi_e}. \label{eq_parallelfric}
\end{equation}
For simplicity a scalar friction is shown here, but these results apply equally well to each component of the friction tensor.
The parallel addition may be rationalised as follows.
When a large particle is forced to move through a sea of smaller particles, it can dissipate energy through two parallel channels: 1) by dragging itself
through this sea of small particles, resulting in more large-small collisions, or 2) by setting up a flow field in the solvent, at the expense of viscous
dissipation in the solvent but with the advantage that the sea of gas particles is co-moving near its surface, resulting in less large-small collisions.

In a real (experimental) colloidal suspension both the solvent density and the range of the colloid-solvent interaction are larger than simulated here,
which is why at any given time many more solvent molecules are interacting with each colloid.
This leads to an Enskog friction which is orders of magnitude larger than the hydrodynamic friction (it is important to note that for a hard sphere in
a point particle fluid the Enskog friction increases faster than the hydrodynamic friction with increasing sphere diameter, $\xi_e \propto D^2$ and
$\xi_h \propto D$).
Hence for real mesoscopic particles the total friction $\xi$ is practically equal to the hydrodynamic friction $\xi_h$.
In our simulations the Enskog friction on one sphere is only approximately 4 times larger than the hydrodynamic friction \cite{Padding06}.

In our simulations, we can easily determine the Enskog friction. In many ways the hard-colloid and SRD fluid system is like a hard sphere system, in which case
the general structural features of the force autocorrelation (or time dependent friction) are (a) an initial delta function contribution whose integral
is the Enskog friction and (b) a slower contribution associated with correlated collisions and collective effects, which is negative
for low and intermediate particle densities \cite{Hynes77}.
The Enskog friction can therefore be read off as the peak value in the running integral of Eq.~(\ref{eq_fricintegral}),
$\Xi_{\alpha \beta}(t) = (1/k_BT) \int_0^t \dif \tau \left\langle F^c_{\alpha}(t_0 + \tau) F^c_{\beta}(t_0) \right\rangle_{t_0}$, at very short times $t$.
Note that for continuous interactions there is no clear-cut separation as described here \cite{Hynes77}.
Applying the above procedure to the case of a sphere yields an Enskog friction which is in good agreement with theoretical prediction, as
we will show in section \ref{sec_sphere}. Eq.~(\ref{eq_parallelfric}) is then inverted to $\xi_h = 1/(1/\xi - 1/\xi_e)$ to obtain the
hydrodynamic friction coefficient which would be measured for a particle with orders of magnitude higher Enskog friction.

\subsection{Choice of system coordinates}

\begin{figure}[tb]
   \begin{center}
     \scalebox{0.4}
     {\includegraphics{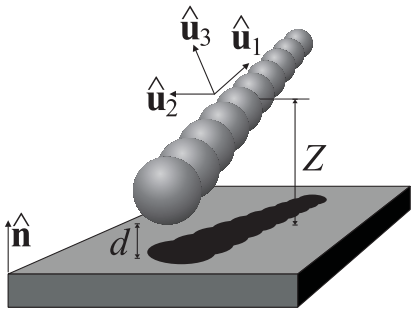}}\\
   \end{center}
  \caption{
    \label{geometry}
     Rod and wall geometry. $Z$ is the height of the rod centre-of-mass above the wall, $d$ is the closest distance between the rod surface and the wall,
     which has its normal in the $\nn$ direction.
     The unit vector $\uu_1$ is along the long axis of the rod, $\uu_2$ is perpendicular to the rod and pointing parallel to the wall surface, and $\uu_3$
     is perpendicular to the previous two. The ange $\theta$ is defined through $\nn \cdot \uu_1 = \cos \theta$. 
}
\end{figure}

Looking at Fig.~\ref{geometry} it is clear that the cartesian coordinates $x$, $y$ and $z$ may not be the most natural coordinates to use when 
considering the friction on a rod. In this section we will introduce coordinates which are better adapted to the symmetry of the problem.
We will show this for the case of the translational friction, but similar results will apply to the rotational friction.

By symmetry, given the orientation of the rod in the $xz$-plane, the components $\Xi_{xy}$ and $\Xi_{yz}$ must be zero (and hence also 
$\Xi_{yx}$ and $\Xi_{zy}$). The remaining 4 independent components are the diagonals and the $xz$-component (which is
equal to the $zx$-component). Their relative importance in general depends on the orientation of the rod.
In order to simplify this dependence as much as possible we transform the friction tensors to a
rod-based orthogonal coordinate system, defined as follows (see Fig.~\ref{geometry}):
\begin{eqnarray}
\uu_1 &=& \uu \\
\uu_2 &=& \frac{\nn \times \uu}{\left| \nn \times \uu \right| } \\
\uu_3 &=& \frac{\uu \times \left( \nn \times \uu \right)}{\left| \nn \times \uu \right| }
\end{eqnarray}
In short, $\uu_1$ is along the rod, $\uu_2$ is perpendicular to the rod but parallel to the wall, and $\uu_3$ is perpendicular to the previous two.
If $\theta$ is the angle between the long axis of the rod and the wall normal ($\nn \cdot \uu = \ct$), with $\theta \in [0,\pi/2]$, 
then the transformation matrix $\mathbf{U}$ between the cartesian frame and this new coordinate system is given by
\begin{equation}
\mathbf{U} = \left[
\begin{array}{ccc}
	\st & 0 & -\ct \\
	0 & 1 & 0 \\
	\ct & 0 & \st 
\end{array}
\right].
\end{equation}
With this we can calculate the transformed friction tensor as
\begin{equation}
\tilde{\mathbf{\Xi}} = \mathbf{U}^T \mathbf{\Xi} \mathbf{U} = \left[
\begin{array}{ccc}
	\xi_{||} & 0 & \xi' \\
	0 & \xi_{\perp 1} & 0 \\
	\xi' & 0 & \xi_{\perp 2} 
\end{array}
\right],
\end{equation}
where the friction components are given by
\begin{eqnarray}
\xi_{||} &=& \sin^2 \theta \ \Xi_{xx} + 2 \st \ct \ \Xi_{xz} + \cos^2 \theta \ \Xi_{zz} \\
\xi_{\perp 1} &=& \Xi_{yy} \\
\xi_{\perp 2} &=& \cos^2 \theta \ \Xi_{xx} - 2 \st \ct \ \Xi_{xz} + \sin^2 \theta \ \Xi_{zz} \\
\xi' &=& \st \ct \left( \Xi_{zz} - \Xi_{xx} \right) + \left( \sin^2 \theta - \cos^2 \theta \right) \Xi_{xz} \nonumber \\
\end{eqnarray}
It will turn out that the mixing term $\xi'$ is small relative to the other terms.

\section{Validation: friction on a sphere near a wall}
\label{sec_sphere}

To validate our method, we first determine the friction on a single sphere as a function of its height $z$ above a wall,
and compare with known theoretical expressions.

\begin{figure}[tb]
  \begin{center}
     \scalebox{0.4}{\includegraphics{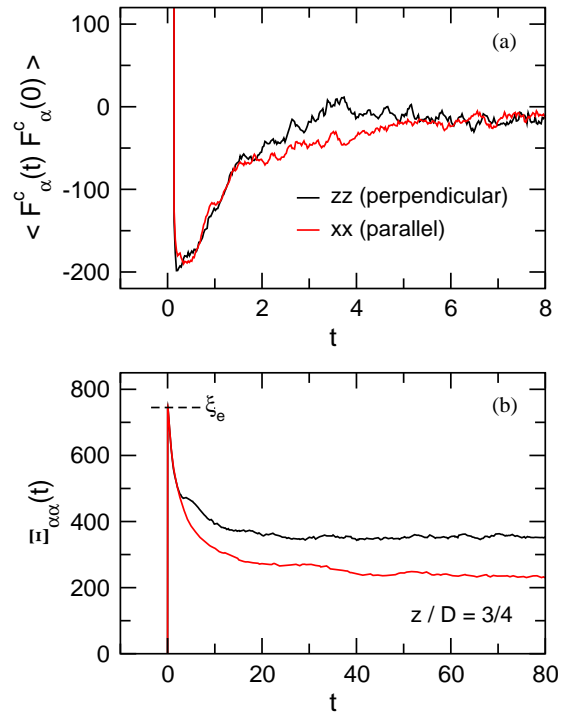}}\\
   \end{center}
  \caption{
    \label{frictionintegral_sphere}
     (a) Autocorrelation of the constraint force needed to keep a sphere at fixed position $Z = 3/4 D$ above a wall. Forces perpendicular
     to the wall ($zz$) are shown in black; forces parallel to the wall ($xx$) in red.
     (b) Running integral of the data in (a) (divided by $k_BT$). The peak value at short times yields the Enskog friction (dashed line) $\xi_e$,
     which determines the short time self-diffusion of a colloid in a Brownian bath without hydrodynamic interactions. The total friction $\xi$ is estimated
     from the long time limit of the running integral. The hydrodynamic friction $\xi_h$ is obtained by applying Eq.~(\ref{eq_parallelfric}).
}
\end{figure}
We have determined the constraint force autocorrelations on a sphere of diameter $D = 8a_0$ for a series of heights ranging from
$z/D = 0.6$ to 5.0 in a cubic box of size $L = 10 D$ along each axis. An example for
$z/D = 0.75$ (i.e. with a gap width of $0.25 D$ between the wall and the bottom of the sphere) is given in
Fig.~\ref{frictionintegral_sphere}(a) where we show both the perpendicular ($zz$) and parallel ($xx$) components.
The structural features of the force autocorrelations are, as
expected \cite{Hynes77}, an initial delta function contribution, whose integral is the Enskog friction, and a slower contribution associated
with correlated collisions and collective effects, which is negative for our relatively low particle density. 

The running integral of the constraint force autocorrelation (divided by $k_BT$) is shown in Fig.~\ref{frictionintegral_sphere}(b).
The peak value at short times measures $0.74 \times 10^3$ in simulation units, which is in good agreement with the expected Enskog
friction of $\xi_e = 0.69 \times 10^3$. We note that the measured peak value of $0.74 \times 10^3$ was found consistently for all distances
between sphere and wall. This confirms that the value of the Enskog friction is a local effect, unaffected by the geometry of the surroundings
of the sphere.

After the Enskog peak, the running integrals converge slowly to their final values, in the particular case shown
$\Xi_{zz} = 0.35 \times 10^3$ and $\Xi_{xx} = 0.23 \times 10^3$, but with other values for other distances between sphere and wall.
Using the measured Enskog friction, we then calculate the perpendicular and parallel hydrodynamic frictions.
For large values of $z/D$ these hydrodynamic frictions converge to a value $\xi^\infty = 230 \pm 20$, which is in good agreement with
the expected value of $\xi^{theor} = 3\pi \eta D / (1-1.45D/L) = 220$, where the factor between brackets takes into account finite system
size effects \cite{Padding06,ZickHomsy,Duenweg}.

\begin{figure}[tb]
   \begin{center}
     \scalebox{0.4}
     {\includegraphics{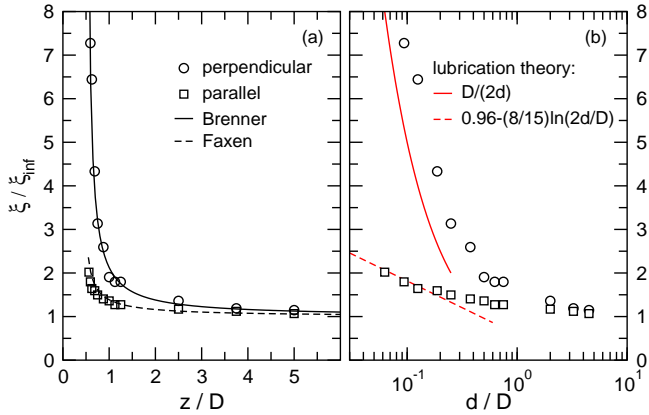}}\\
   \end{center}
  \caption{
    \label{spherefriction}
     (a) Hydrodynamic friction enhancement due to the presence of a wall as a function of normalised
     height $Z/D$ of the sphere's centre above the wall.
     Simulation results (circles and squares) are compared with theoretical expressions (black lines)
     for the friction for perpendicular motion towards and away from the wall, Eq.~(\ref{eq_perp})
     \cite{Brenner}, and for motion parallel to the wall, Eq.~(\ref{eq_paral}) \cite{Faxen}.
     (b) Hydrodynamic friction enhancement as a function of normalised closest distance $d/D$,
     with $d=Z-D/2$,
     on a semi-logarithmic scale. The simulation results (circles and squares) are compared with
     lubrication theory asymptotes \cite{Brenner,Goldman} which are valid for $d \ll D$ (red lines).
}
\end{figure}
In Fig.~\ref{spherefriction}(a) we plot the resulting perpendicular (circles) and parallel (squares) hydrodynamic frictions,
normalised by $\xi^\infty$, as a function of normalised height $z/D$ of the sphere's centre above the wall.
Clearly, the friction is enhanced greatly as the sphere comes nearer to the wall. A theoretical derivation for the perpendicular friction
enhancement $\lambda_{\perp}$ was given in 1961 by Brenner \cite{Brenner}, with the result
\begin{eqnarray}
&& \lambda_{\perp}(z) = \frac43 \sinh \alpha \sum_{n=1}^{\infty} \frac{n(n+1)}{(2n-1)(2n+3)} \times \nonumber \\
&& \left[ \frac{2\sinh ((2n+1)\alpha) + (2n+1)\sinh(2 \alpha)}{\left[ 2 \sinh ((n+1/2) \alpha) \right]^2 - \left[ (2n+1) \sinh \alpha \right]^2} -1 \right], \nonumber \\ \label{eq_perp}
\end{eqnarray}
where $\alpha = \cosh^{-1}(2z/D)$. Fig.~\ref{spherefriction}(a) shows this theoretical result as a solid line.
There is no exact analytical expression for the parallel friction enhancement $\lambda_{||}$. A commonly applied approximation due to Fax\'en \cite{Faxen},
which deviates less than 10\% from the result of precise numerical calculations for $z/D > 0.52$ and gives essentially the same results for $z/D > 0.7$ \cite{Goldman},
is the following:
\begin{eqnarray}
\lambda_{||}(z) &=& \left[ 1 - \frac{9}{32}\frac{D}{z} + \frac{1}{64}\left( \frac{D}{z} \right)^3 \right. \nonumber \\
&& \left. - \frac{45}{4096} \left( \frac{D}{z} \right)^4 
-\frac{1}{512} \left( \frac{D}{z} \right)^5 \right]^{-1}. \label{eq_paral}
\end{eqnarray}
Fig.~\ref{spherefriction}(a) shows this expression as a dashed line.

When the closest distance $d = Z-D/2$ is much
smaller than the sphere diameter, i.e. when $d \ll D$, the Stokes equations can be solved asymptotically, leading
to so-called lubrication forces \cite{Brenner,Goldman}. For the perpendicular friction enhancement 
the lubrication prediction diverges as the inverse closest distance \cite{Brenner}:
\begin{equation}
\lim_{d/D \to 0} \lambda_{\perp} = \frac{D}{2d}. \label{eq_lubricationperp}
\end{equation}
For the parallel friction enhancement the lubrication prediction diverges logarithmically \cite{Goldman},
\begin{equation}
\lim_{d/D \to 0} \lambda_{||} \approx 0.9588 - \frac{8}{15} \ln \frac{2d}{D},
\end{equation}
where the constant 0.9588 has been determined by fitting to precise numerical calculations \cite{Goldman}.
An interesting question is whether our simulations are able to reproduce these asymptotes.
Lubrication forces are in essence a hydrodynamic effect and the SRD method in principle resolves
fully the hydrodynamics, at least down to the scale of a cell size $a_0$. The resolution is even better than
this because of the averaging effect of the applied random grid shift \cite{Ihle}.

Fig.~\ref{spherefriction}(b) shows the friction enhancements measured in the simulations and the lubrication
predictions against the logarithm of $d/D$, thus emphasising the behaviour at small distances.
We observe that our method predicts a perpendicular friction which approaches the lubrication
prediction (solid red line) from above, but does not yet reach this limit within the range of distances studied.
The approach is quite slow, which is in agreement with the exact expression Eq.~(\ref{eq_perp}).
The parallel friction approaches the lubrication prediction (dashed red line) from above too, but here
the lubrication limit is reached already for $d/D < 0.1$, in agreement with observations by
Goldman \textit{et al.} \cite{Goldman}.

The good agreement of our results with the theoretical predictions for a sphere,
both at small and large distances, gives us confidence that
the same method may also be applied to the case of a rod for which no theoretical predictions are known.
In the remainder of this paper we will focus on closest distances in the range of $D/8$ to
$10 D$. We will show that in this range of distances the wall-induced friction enhancement can be approximately described by additive contributions scaling linearly with $D/d$. We emphasise that this scaling is not exact,
but serves to represent our measurements in a compact functional form which may be useful for future simulations.
The similarity of the $d^{-1}$ scaling with Eq.~(\ref{eq_lubricationperp}) is probably
coincidental because lubrication theory generally is not valid at such large distances.

\section{Friction on a rod near a wall with $L/D = 10$}
\label{sec_rod}

\subsection{Translational friction}

\begin{figure}[tb]
\begin{center}
     \scalebox{0.4}{\includegraphics{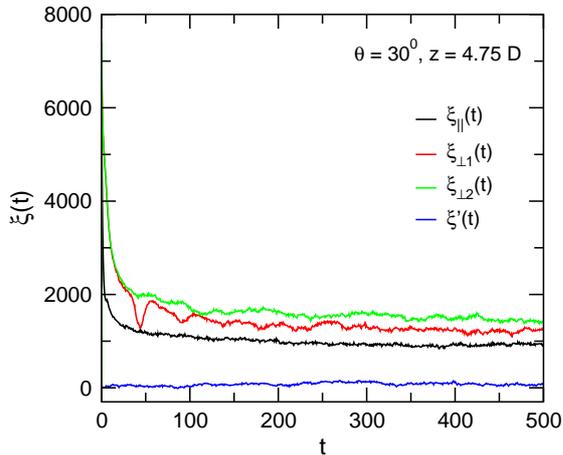}}\\
   \end{center}
  \caption{
    \label{frictionintegral_rod}
     Running integral of the autocorrelation of the constraint force needed to keep a rod at fixed position $Z = 4.75 D$ and orientation $\theta = 30^0$.
     The diagonal components represent the magnitude of the friction anti-parallel to the direction of motion, for motion along $\uu_1$ ($\xi_{||}$),
     $\uu_2$ ($\xi_{\perp 1}$), and $\uu_3$ ($\xi_{\perp 2}$). The mixing term  $\xi'$ represents friction along the $\uu_1$ direction for motion along the
     $\uu_3$ direction (and vice-versa). The mixing term is always much smaller than the diagonal components.
}
\end{figure}
A typical example of the running integral of the constraint force autocorrelation for a rod near a wall is given in Fig.~\ref{frictionintegral_rod}.
The diagonal components
$\xi_{||}$, $\xi_{\perp 1}$, and $\xi_{\perp 2}$ represent the magnitude of the friction anti-parallel to the direction of motion, for motion
along $\uu_1$, $\uu_2$, and $\uu_3$, respectively (note that in our simulations we do not really move the particles). 
The mixing term  $\xi'$ represents friction along the $\uu_1$ direction for motion along the
$\uu_3$ direction (and vice-versa).
The mixing term is always found to be at least one order of magnitude smaller than the three diagonal components, so to a first approximation may be neglected. The fact that the mixing term is always much smaller than the diagonal components shows that $\uu_1$, $\uu_2$ and $\uu_3$ are indeed close
to the principal axes of the friction tensor.
Moreover, the convergence of the friction data was confirmed by performing duplo runs for most systems, yielding
identical results (including, for example, the oscillation visible in $\xi_{\perp 1}(t)$ near $t = 50t_0$).

\begin{figure}[tb]
\begin{center}
     \scalebox{0.45}
     {\includegraphics{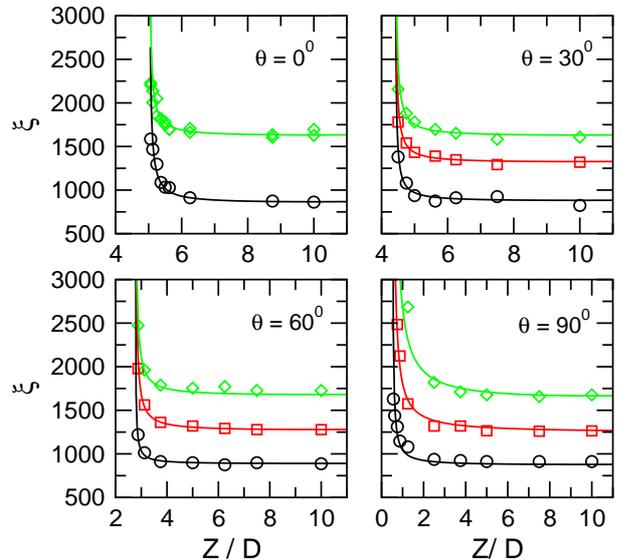}}\\
   \end{center}
  \caption{
    \label{translationalfriction}
     Hydrodynamic translational friction components $\xi_{||}$ (black circles), $\xi_{\perp 1}$ (red squares) and $\xi_{\perp 2}$ (green diamonds)
     as a function of rod height $Z$, for four different angles $\theta$. Solid lines are fits using Eq.~(\ref{eq_transfric}).
}
\end{figure}
The Enskog friction was again determined from the peak value of the running integral at short times and the total friction from the limiting value at
large times. The resulting hydrodynamic frictions are presented in Fig.~\ref{translationalfriction} (symbols) as a function of distance between rod and
wall for four values of the angle $\theta$.
We find that a reasonable approximation (within the range of distances studied) for the translational friction components is to treat the wall effect as additional
to the bulk friction, with a dominant dependence on the inverse smallest distance between the surface of the
rod and the wall and an angle-dependent prefactor. The smallest distance is defined as follows: a shish-kebab rod
consisting of $L/D$ spheres, with its centre-of-mass at height $z = Z$, making an angle
$\theta$ with the wall normal, will have a smallest distance $d_1$ with the wall at $z=0$ given by
\begin{equation}
d_1 = Z - \left[ \frac{L/D - 1}{2} \ct + \frac12 \right] D.
\end{equation}
Another wall is present at $z = L_z$, with a smallest distance $d_2$ to the surface of the rod given by
$d_2 = L_z + d_1 - 2Z$. Within our approximation, the translational friction components may be expressed as
\begin{equation}
\xi_{\alpha \beta} \approx \xi_{\alpha \beta}^{\infty} \left[ 1 + A_{\alpha \beta}(\theta) \left( \frac{D}{d_1} + \frac{D}{d_2} \right) \right], \label{eq_transfric}
\end{equation}
where $\xi_{\alpha \beta}$ can be any of the $\xi_{||}$, $\xi_{\perp 1}$, or $\xi_{\perp 2}$ components.
The fits are presented in Fig.~\ref{translationalfriction} as solid lines. The $z=\infty$ values and prefactors $A$ are estimated by a least-squares-fit.
Note that the prefactors $A$ may in principle also depend on the aspect ratio $p = L/D$; our results apply to the case $p = 10$ only.
We find the following results for the bulk values (in our units): $\xi_{||}^\infty = 860 \pm 20$, $\xi_{\perp 1}^\infty = 1250 \pm 50$
and $\xi_{\perp 2}^\infty = 1500 \pm 80$, independent of the particular value of $\theta$. 
These values may be compared to the approximate theoretical predictions \cite{Tirado84} $\xi_{||}^{theor} = 2\pi \eta L/(\ln p -0.207+0.908/p) = 575$ and
$\xi_{\perp}^{theor} = 4\pi \eta L/(\ln p + 0.839 + 0.185/p) = 800$ valid for a cylindrical rod in an infinite solvent bath.
The frictions we measure are higher because of unavoidable self-interactions between the rod and its periodic images which overall tend
to increase the friction. In other words, even when the walls are infinitely far apart (in the $z$-direction), the periodic boundaries in the other
two directions still cause friction enhancements of each of the friction components. We find that the self-interactions enhance the component
$\xi_{||}^\infty$ by the same amount, independent of the actual rod angle $\theta$. Similarly the values for $\xi_{\perp 1}^\infty$ and $\xi_{\perp 2}^\infty$
are consistently the same for all rod angles.

\begin{figure}[bt]
\begin{center}
     \scalebox{0.4}
     {\includegraphics{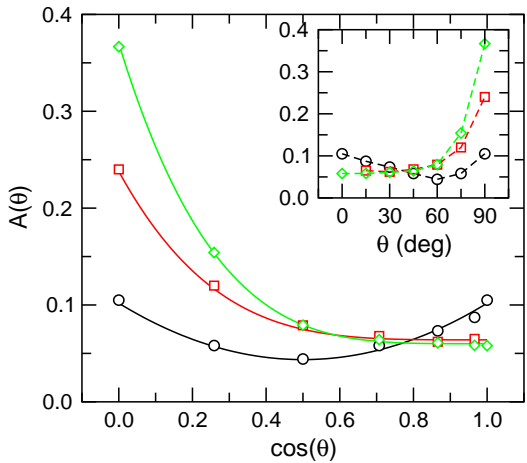}}\\
   \end{center}
  \caption{
    \label{transprefactorscaling}
     Prefactors $A_{||}(\theta)$ (black circles), $A_{\perp 1}(\theta)$ (red squares), and $A_{\perp 2}(\theta)$ (green diamonds), as obtained from fits
     to translational friction coefficients using Eq.~(\ref{eq_transfric}), versus rod angle $\theta$ (inset) or $\cos \theta$ (main plot).
     Solid lines in the main plot are fits using Eqs.~(\ref{eq_prefactortrans1}) - (\ref{eq_prefactortrans3}).
}
\end{figure}
In Fig.~\ref{transprefactorscaling} (inset) we present the prefactors $A$, as obtained from the fits, as a function of rod angle $\theta$.
In the absence of theoretical predictions we have tried several fit functions. Good single powerlaw fits can be made when the prefactors are 
plotted against $\cos \theta$, see Fig.~\ref{transprefactorscaling} (main plot), resulting in the following fit functions:
\begin{eqnarray}
A_{||}(\theta) &=& A^0_{||} + B_{||} \left( 1/2 - \ct \right)^2 \label{eq_prefactortrans1} \\
A_{\perp 1}(\theta) &=& A^0_{\perp 1} + B_{\perp 1} \left( 1 - \ct \right)^4 \\
A_{\perp 2}(\theta) &=& A^0_{\perp 2} + B_{\perp 2} \left( 1 - \ct \right)^4, \label{eq_prefactortrans3}
\end{eqnarray}  
with $A^0_{||} = 0.044 \pm 0.003$, $B_{||} = 0.23 \pm 0.02$,
$A^0_{\perp 1} = 0.064 \pm 0.005$, $B_{\perp 1} = 0.17 \pm 0.02$, $A^0_{\perp 2} = 0.060 \pm 0.004$, and $B_{\perp 2} = 0.31 \pm 0.02$.

In summary, if only one wall is present near a rod of $L/D = 10$ with closest distance $d$, the translational friction components are approximately given by
\begin{eqnarray}
\xi_{||} &\approx & \xi_{||}^{\infty} \left\{ 1 + \left[0.044 + 0.23 \left( 1/2 - \ct \right)^2 \right] \frac{D}{d} \right\} \label{eq_trans1} \\
\xi_{\perp 1} &\approx & \xi_{\perp}^{\infty} \left\{ 1 + \left[ 0.064 + 0.17 \left( 1 - \ct \right)^4 \right] \frac{D}{d} \right\} \\
\xi_{\perp 2} &\approx & \xi_{\perp}^{\infty} \left\{ 1 + \left[ 0.060 + 0.31 \left( 1 - \ct \right)^4 \right] \frac{D}{d} \right\} \label{eq_trans3}
\end{eqnarray}

\subsection{Rotational friction}

\begin{figure}[tb]
\begin{center}
     \scalebox{0.45}
     {\includegraphics{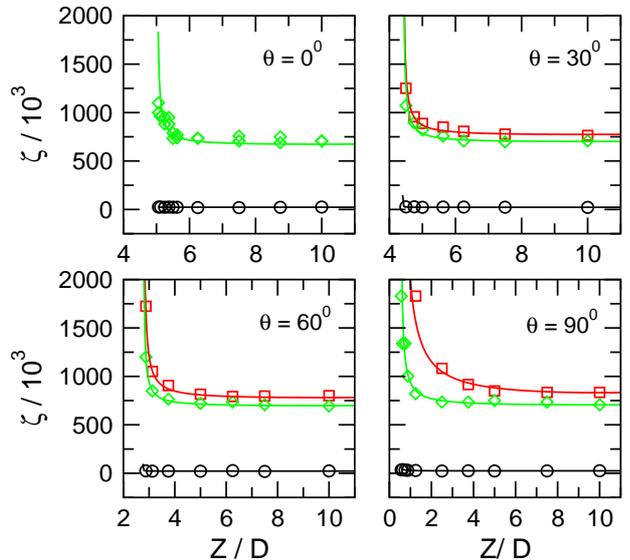}}\\
   \end{center}
  \caption{
    \label{rotationalfriction}
     Hydrodynamic rotational friction components $\zeta_{||}$ (black circles), $\zeta_{\perp 1}$ (red squares) and $\zeta_{\perp 2}$ (green diamonds)
     as a function of rod height $Z$, for four different angles $\theta$. Solid lines are fits using Eq.~(\ref{eq_rotfric}).
}
\end{figure}
Figure \ref{rotationalfriction} (symbols) shows the hydrodynamic rotational friction coefficients as a function of distance between rod and wall
for four different values of the angle $\theta$. The component $\zeta_{||}$ (circles) represents the rotational friction for rotation around the long
($\uu_1$) axis, $\zeta_{\perp 1}$ represents the rotational friction for rotation around the $\uu_2$ axis, and $\zeta_{\perp 2}$ the rotational friction
for rotation around the $\uu_3$ axis. The mixing term $\zeta'$ was again found to be at least one order of magnitude smaller, and is therefore
neglected in the following analysis.

Similarly to the translational friction, we have treated the wall effect as additive to the bulk rotational friction, again with a dominant
inverse dependence on the smallest distance $d$. Denoting rotational friction components with $\zeta$, these may be expressed as
\begin{equation}
\zeta_{\alpha \beta} \approx \zeta_{\alpha \beta}^{\infty} \left[ 1 + C_{\alpha \beta}(\theta) \left( \frac{D}{d_1} + \frac{D}{d_2} \right) \right], \label{eq_rotfric}
\end{equation}
where $\zeta_{\alpha \beta}$ can be any of the $\zeta_{||}$, $\zeta_{\perp 1}$, or $\zeta_{\perp 2}$ components. The fits are
represented in Fig.~\ref{rotationalfriction} as solid lines.
Again note that the prefactors $C$ may also depend on the aspect ratio $p = L/D$. We find the following values for the bulk
values (in our units): $\zeta_{||}^\infty = (0.22 \pm 0.01) \cdot 10^5$, $\zeta_{\perp 1} = (7.4 \pm 0.1) \cdot 10^5$ and $\zeta_{\perp 2} = (6.7 \pm 0.2) \cdot 10^5$.
Theoretically, the expression for a cylindrical rod of aspect ratio $p = 10$ for the latter two reads \cite{Tirado84}:
$\zeta_{\perp}^{theor} = \pi \eta L^3/[3 (\ln p -0.662 + 0.917/p)] = 7.7 \cdot 10^5$.
We consider the rather good agreement with our results to be somewhat fortituous: the theoretical results have been derived for a
cylinder of length $L$ and diameter $D$ in an infinite bath, whereas we have simulated a succession of spheres in a finite bath.
Because forces on the the extremes of a rod have the most important contribution to the torque, the magnitude of the rotational friction
on a rod is much more sensitive to the shape of its extremes than the translational friction.
The rounded extremes of our model would correspond effectively to a cylinder of smaller length (for example for a cylinder with $L = 9D$
a rotational friction of $\zeta_{\perp} = 6.0 \cdot 10^5$ would be predicted).

The rotational friction around the long axis is less than $3 \%$ of those around the two perpendicular axes, and relatively it remains much smaller
also when the distance to a wall becomes very small.

\begin{figure}[tb]
\begin{center}
     \scalebox{0.4}
     {\includegraphics{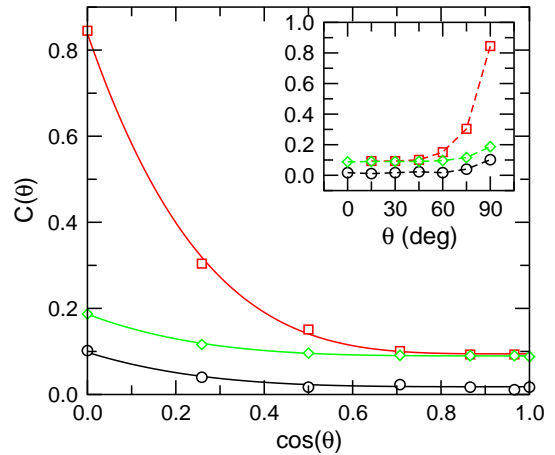}}\\
   \end{center}
  \caption{
    \label{rotprefactorscaling}
     Prefactors $C_{||}(\theta)$ (black circles), $C_{\perp 1}(\theta)$ (red squares), and $C_{\perp 2}(\theta)$ (green diamonds), as obtained from fits
     to rotational friction coefficients using Eq.~(\ref{eq_rotfric}), versus rod angle $\theta$ (inset) or $\cos \theta$ (main plot).
     Solid lines in the main plot are fits using Eqs.~(\ref{eq_prefactorrot1}) - (\ref{eq_prefactorrot3}).
}
\end{figure}
In Fig.~\ref{rotprefactorscaling} (inset) we present the prefactors $C$, as obtained from the fits, as a function of rod angle $\theta$.
Good single powerlaw fits can again be made with our measurements when they are plotted against $\cos \theta$ (main plot), resulting in the following fit functions:
\begin{eqnarray}
C_{||}(\theta) &=& C^0_{||} + E_{||} \left( 1 - \ct \right)^4 \label{eq_prefactorrot1} \\
C_{\perp 1}(\theta) &=& C^0_{\perp 1} + E_{\perp 1} \left( 1 - \ct \right)^4 \\
C_{\perp 2}(\theta) &=& C^0_{\perp 2} + E_{\perp 2} \left( 1 - \ct \right)^4 \label{eq_prefactorrot3}
\end{eqnarray}  
with $C^0_{||} = 0.036 \pm 0.002$, $E_{||} = 0.080 \pm 0.006$, $C^0_{\perp 1} = 0.094 \pm 0.004$, $E_{\perp 1} = 0.74 \pm 0.04$, $C^0_{\perp 2} = 0.090 \pm 0.004$,
and $E_{\perp 2} = 0.097 \pm 0.005$.

In summary, if one wall is present near a rod of $L/D = 10$ with closest distance $d$, the rotational friction components are approximately given by
\begin{eqnarray}
\zeta_{||} &\approx & \zeta_{||}^{\infty} \left\{ 1 + \left[ 0.036 + 0.08 \left( 1 - \ct \right)^4 \right] \frac{D}{d} \right\} \label{eq_rot1} \\
\zeta_{\perp 1} &\approx & \zeta_{\perp}^{\infty} \left\{ 1 + \left[ 0.094 + 0.74 \left( 1 - \ct \right)^4 \right] \frac{D}{d} \right\}  \\
\zeta_{\perp 2} &\approx & \zeta_{\perp}^{\infty} \left\{ 1 + \left[ 0.090 + 0.097 \left( 1 - \ct \right)^4 \right] \frac{D}{d} \right\} \label{eq_rot2}
\end{eqnarray}

\section{Conclusion}
\label{sec_concl}

We have shown that Stochastic Rotation Dynamics simulations can be used to measure the hydrodynamic friction on an object by constraining its
position and orientation and analysing the time correlation of the constraint force.

In this work we have applied the method to a rod of aspect ratio $L/D = 10$ near a wall.
The main result is summarised in Eqs. (\ref{eq_trans1}) - (\ref{eq_trans3}) and (\ref{eq_rot1}) - (\ref{eq_rot2}). Reasonably good fits of both the translational
and rotational friction could be made with the inverse of the closest distance $d$ between the rod and wall, at least in the range $d \in [D/8,L]$.

We have found that
for a noticeable friction
increase the closest distance between rod and wall needs to be on the order of the rod diameter. Also, in agreement with common sense, the friction increase
is strongest when the rod lies parallel to the wall ($\ct = 0$). For translations, the friction component $\xi_{\perp 2}$ is the largest, as this corresponds
(at least partially) to motion to and from the wall. For rotations, the friction component $\zeta_{\perp 1}$ is the largest, as this corresponds to motion to
and from the wall of the extremities of the rod.

In anticipation of an accurate theoretical treatment of this system,
we have fitted the angular dependence of the friction components and found good fits in most cases with $(1 - \ct)^4$.
We do not have a motivation for this functional form except that, intuitively, the friction increase must be
relatively larger when a larger area of the rod is exposed close to the wall. Hence an increasing function
of angle $\theta$ is expected. The only exception seems to be the angular dependence of the parallel
translational friction $\xi_{||}$, for which the smallest friction increase occurs at an intermediate angle of
60 degrees (see the inset of Fig.~\ref{transprefactorscaling}). Because we cannot give a full physical motivation,
the scalings we have presented here are possibly not exact. However, they do serve to represent our measurements
in a compact form which may be useful for future simulations. Such simulations are planned for the near future.

More generally, the technique presented in this paper offers the posibility to determine
with reasonable precision the hydrodynamic frictions
on complex objects, possibly with nearby complex boundaries, under circumstances where theoretical calculations are too difficult to be performed.

\section*{Acknowledgements}

We thank Peter Lang and Jan Dhont for useful discussions. We thank the NoE `SoftComp' and NMP SMALL `Nanodirect' for financial support. JTP thanks the
Netherlands Organisation for Scientific Research (NWO) for additional financial support.

\end{document}